\documentclass[
aps,
nofootinbib,
superscriptaddress,
tightenlines,
notitlepage,
twocolumn,
showpacs,
floatfix,
]{revtex4-1}
\usepackage{amssymb,amsmath,bm,tensor,braket}
\usepackage[colorlinks]{hyperref}

\usepackage{graphicx}
\usepackage{dcolumn}
\usepackage{bm}
\usepackage{xcolor}

\newcommand{\PlanckMass}{M_{\rm Pl}}

\begin{document}


\title{ Cosmological aspects of cubic Galileon massive gravity}

\author{Shahabeddin M. Aslmarand}
 \email{smostafanazh2016@fau.edu}
\affiliation{%
Department of Physics, Florida Atlantic University, Boca Raton, FL 33431, USA
}%

\author{Amin Rezaei Akbarieh}
 \email{am.rezaei@tabrizu.ac.ir}
\affiliation{
 Faculty of Physics, University of Tabriz, Tabriz 51666-16471, Iran
}%

\author{Yousef Izadi}
 \email{yousef_izadi@uml.edu}
\affiliation{%
Department of Physics and Applied Physics, University of Massachusetts, Lowell, MA 01854, USA
}%

\author{Sobhan Kazempour}
 \email{s.kazempour@tabrizu.ac.ir}
\affiliation{%
Faculty of Physics, University of Tabriz, Tabriz 51666-16471, Iran
}%

\author{Lijing Shao}
 \email{lshao@pku.edu.cn}
\affiliation{%
Kavli Institute for Astronomy and Astrophysics, Peking University, Beijing 100871, China\\
National Astronomical Observatories, Chinese Academy of Sciences, Beijing 100012, China 
}%

\date{\today}

\begin{abstract}
Cubic Galileon massive gravity is a development of de Rham-Gabadadze-Tolley
(dRGT) massive gravity theory in which the space of the Stueckelberg field is
broken.  We consider the cubic Galileon term as a scalar field coupled to the
graviton field. We present a detailed study of the cosmological aspects of this
theory of gravity.  We analyze self-accelerating solutions of the background
equations of motion to explain the accelerated expansion of the Universe.
Exploiting the latest Union2 Type Ia Supernovae (SNIa) dataset, which consists
of 557 SNIa, we show that cubic Galileon massive gravity theory is consistent
with the observations. We also examine the tensor perturbations within the
framework of this model and find an expression for the dispersion relation of
gravitational waves, and show that it is consistent with the observational
results
\end{abstract}

\maketitle


\section{Introduction} \label{sec:intro}

Although Einstein's theory of general relativity has been very successful in
describing gravity at low energies and explaining observations in Solar system
scales, this theory faces fundamental problems in cosmology
\cite{Will:2001mx,Reynaud:2008yd,Everitt:2011hp,Berti:2015itd,Arkani-Hamed:1998jmv}.
General relativity cannot explain the origin of accelerated expansion of the
Universe and the cosmological constant problem
\cite{Weinberg:1988cp,Peebles:2002gy,SupernovaSearchTeam:1998fmf}. There are
strong observational evidences such as cosmic background radiation (CMB)
\cite{Planck:2015fie}, type Ia supernovae \cite{Phillips:1993ng} and baryon acoustic
oscillations \cite{Beutler:2011hx} which confirm the existence of these
problems. The Universe can be described by the standard models of particle
physics and cosmology in microscopic and large cosmological scales,
respectively. It is noticeable that many physicists would like to unify these
two models into a single comprehensive theory.

One tentative approach for solving the problems in gravity and cosmology is
studying a spin-2 massive graviton as a propagator of gravity. Many models have
been developed to describe graviton and its interactions. Still, the main
concern is finding a theory that would be stable and consistent with
observations. It is interesting to note that the first attempt to explain a
massive spin-2 particle began by Fierz and Puali in 1939; they proposed a linear
action for a massive spin-2 particle (i.e., massive graviton) in a flat
space-time \cite{Fierz:1939ix}.  Vainshtein proposed an idea to solve the vDVZ
discontinuity, which van Dam, Veltman, and Zakharov found in the limit of the
graviton mass $m_g \rightarrow 0$ in the linear Fierz-Pauli action, see
Refs.~\cite{Vainshtein:1972sx,vanDam:1970vg,Zakharov:1970cc}. Vainshtein argued
that to aviod the vDVZ discontinuity, Fierz and Pauli's theory should be
nonlinear instead of linear. Moreover, in 1979, Boulware and Deser claimed that
the nonlinear theory of Fierz and Pauli has a ghost instability which was later
called the Boulware-Deser ghost \cite{Boulware:1972yco}.

Eventually, in 2010 de Rham, Gabadadze, and Tolley showed that it is possible to
construct a ghost-free nonlinear massive gravity in a certain decoupling limit
\cite{deRham:2010ik}. They presented a theory with nonlinear interactions which
explains massive spin-2 field in a flat spacetime, which is known as a
ghost-free dRGT massive gravity theory \cite{deRham:2010kj}.  However, the dRGT
massive gravity theory consists of instabilities in FLRW homogeneous spacetime
\cite{DeFelice:2012mx}. So, this issue has motivated to people to propose
alternative theories
\cite{DAmico:2011eto,Gumrukcuoglu:2012aa,DeFelice:2013awa,Huang:2012pe,Hassan:2011zd,Hinterbichler:2012cn}.
One of the best solutions is using a coupling background to a scalar field which
is known as quasi-dilaton massive gravity theory \cite{DAmico:2012hia}. This is
successful in explaining the accelerated expansion of the Universe. However, due
to the instability of tensor perturbations in this theory, extensions of it have
been introduced to find an improved solution
\cite{DeFelice:2013tsa,Mukohyama:2014rca}.

In this paper, we consider the scalar field to be a cubic Galileon which is coupled to the graviton field. Using the Galileon model, we introduce a scalar field as a candidate for dark energy. Thus, the Galileon model is one of the generalizations of general relativity and can describe the accelerated expansion of the Universe. In this model, the scalar field is invariant under the Galileon transformation $\phi\rightarrow \phi+b_{\mu}x^{\mu}+c$ where $\phi$ is the Galileon field, $b_{\mu}$ and $c$ are constants. The equations of motion obtained from this model are quadratic, and screening occurs via the Vainshtein mechanism. This symmetry was initially proposed in the separation limit of the Dvali-Gabadadze-Porrati (DGP) brane-world model, which is inherited from Poincaré invariance in higher dimensions \cite{ Luty:2003vm,Dvali:2000rv}. This symmetry was later used in a more general scalar field theory in which the equations of motion are quadratic. Although Galileons were originally introduced in the literature of the DGP brane-world model, they also appear in other gravitational models such as massive gravity \cite{deRham:2010ik,deRham:2010kj,Shao:2020fka}.

In the recent literature, the word ``Galileon'' goes far beyond models in which
the action is invariant under a symmetric translational transformation, and it
is in a class of theories that have derivatives of a coupled field and can
establish a consistent cosmological theory. It has been shown in
Refs.~\cite{Kobayashi:2010wa,Chow:2009fm} that Galileons can properly explain
dark energy. Also, the results of Refs.~\cite{Germani:2016gzh,Burrage:2010cu}
show that this model provides an explanation for the origins of the density
perturbations in the inflationary era; hence it could even be an alternative to
the inflationary models.  There has been a tendency towards studying
cosmological aspects and perturbation analysis in massive gravity theories and
their extensions. For instance, investigation of the constraints on
quasi-dilaton massive gravity for finding the bound on graviton mass, see
Ref.~\cite{Martinovic:2019hpo}. Also, dynamical equations, which lead to the
expansion history of the Universe throughout all eras in extended quasi-dilaton
massive gravity, have been studied in Ref.~\cite{Kahniashvili:2014wua}, and in
this paper,  the effective mass of gravitational waves has been found. Moreover,
the new extension of massive gravity theory has been introduced by breaking the
translation symmetry in the Stueckelberg space \cite{Kenna-Allison:2020egn}.
The perturbation analysis and the propagation of gravitational waves have been
studied in the paper as well. Furthermore, some aspects of Galileon in cosmology
and modified gravity have been studied \cite{Nicolis:2008in,deRham:2011by,
Andrews:2013ora}. It is interesting to note that by considering gravitational
radiation in binary pulsars in the context of the cubic Galileon massive
gravity, a new bound of graviton mass has been obtained in
Ref.~\cite{Shao:2020fka}.  Some other valuable studies are in
Refs.~\cite{Gumrukcuoglu:2020utx,Akbarieh:2021vhv,Gumrukcuoglu:2013nza,Guarato:2013gba}.

The outline of this paper is as follows. In Section \ref{model} we introduce the cubic Galilean massive gravity theory. Also, we obtain the background equations of motion and self-accelerating solutions. In Section \ref{C-t} we test the solutions of cubic Galilean massive gravity theory with the latest Union2 type Ia Supernovae (SNIa) dataset, which consists of 557 SNIa. In Section \ref{T-p} we present a perturbation analysis for determining the dispersion relation of gravitational waves in this theory. In Section \ref{Con}, some concluding remarks are given.

\section{The model}\label{model}

In this section, we introduce a model in which the scalar field is chosen from the cubic Galileon model. The action of the theory can be written as
\begin{eqnarray}
\label{1}
\mathcal{S} = \frac{\PlanckMass^{2}}{2}\int d^{4}x\sqrt{-g} \Big\lbrace R[g]-2\Lambda +2m_{g}^{2}(\mathcal{L}_{2}+\alpha_{3}\mathcal{L}_{3}+\alpha_{4}\mathcal{L}_{4})
\nonumber\\-\frac{\omega}{\PlanckMass^{2}}\partial_{\mu}\sigma\partial^{\mu}\sigma (1+\beta\partial_{\mu}\partial^{\mu}\sigma ) \Big\rbrace , \nonumber\\
\end{eqnarray}
where $\PlanckMass$ is the reduced Planck mass, $g$ is the determinant of the metric, and $R$ is the Ricci scalar. It is important to note that the kinetic part of the action in eq. (\ref{1}) is a subclass of a more general action known as the Horndeski action.  There is a linear expression in the action of the cubic Galileon \cite{Zhang:2020qkd} that plays the role of potential. Here, we discard this linear term since the theory is invariant under a global dilation of the spacetime coordinates accompanied by a shift of $\sigma$. $\beta$ is a cubic Galileon parameter; for $\beta=0$ the action in eq. (\ref{1}) reduces to the standard quasi-dilaton action without any potential term. We assume that $m_{g}$ is a constant parameter, so this model can be considered as a generalization of the quasi-dilaton theory. Part of the action that creates the mass for graviton can be expressed as
\begin{eqnarray}
\label{2}
\mathcal{L}_{2}&&=\frac{1}{2}([\mathcal{K}]^{2}-[\mathcal{K}^{2}]),\nonumber\\
\mathcal{L}_{3}&&=\frac{1}{3!}([\mathcal{K}]^{3}-3[\mathcal{K}][\mathcal{K}^{2}]+2[\mathcal{K}^{3}]),\nonumber\\
\mathcal{L}_{4}&&=\frac{1}{4!}([\mathcal{K}]^{2}-6[\mathcal{K}]^{2}[\mathcal{K}^{2}]+3[\mathcal{K}^{2}]^{2}+8[\mathcal{K}][\mathcal{K}^{3}]-6[\mathcal{K}^{4}]),
\end{eqnarray}
where square brackets denote a trace. While these expressions are in a similar form to dRGT theory, in quasi-dilaton case, the building block tensor $\mathcal{K}$ is defined as
\begin{eqnarray}
\label{3}
\mathcal{K}^{\mu}_{\nu} &=& \delta^{\mu}_{\nu} -e^{\frac{\sigma}{\PlanckMass}}(\sqrt{g^{-1}\tilde{g}})^{\mu}_{\nu},
\end{eqnarray}
where $\tilde{g}$ is a non-dynamical fiducial metric. This symmetry rules out a non-trivial potential for $\sigma$. Throughout the paper, we adopt the units $c = \hbar = 1$ in which the reduced Planck mass becomes $\PlanckMass = \frac{1}{\sqrt{8\pi G}}$. Furthermore, we follow the ``mostly plus'' metric signature convention. Some short-cut notations are used to denote the contractions of rank-2 tensors $\mathcal{K}^{\mu}_{\mu}=[\mathcal{K}]$ and $\mathcal{K}^{\mu}_{\nu}\mathcal{K}^{\nu}_{\mu}=[\mathcal{K}^{2}]$, etc. Greek indices run from 0 to 3 while Latin indices from 1 to 3. With the Latin indices we denote contractions in the same way as for the Greek indices $h_{ij}h^{ij} = (h_{ij})^{2}$, $A_{i}A^{i} = (A_{i})^{2}$, etc.\\

For the physical background metric, we use the flat Friedmann-Lema\^itre-Robertson-Walker  (FLRW) ansatz
\begin{eqnarray}
\label{4}
g_{\mu\nu}=-N(t)^{2}dt^{2}+a(t)^{2}\delta_{ij}dx^{i}dx^{j},
\end{eqnarray}
and Minkowski metric is used for non-dynamical fiducial metric denoted by $\tilde{g}_{\mu\nu}$ and expressed as
\begin{eqnarray}\label{5}
\tilde{g}_{\mu\nu}=-f^{\prime}(t)^{2}dt^{2}+\delta_{ij}dx^{i}dx^{j}.
\end{eqnarray}
Notice that $N$ shows the lapse function of the dynamical metric{\color{blue}{,}} and it is similar to a gauge function. The scale factor is represented by $a${\color{blue}{,}} and the dot denotes the derivative with respect to time. In addition, the lapse function N is related to the coordinate time $t$ and the proper time via $\tau d\tau = Ndt$ \cite{Scheel:1994yn,Christodoulakis:2013xha}. Also, $f(t)$ is the Stueckelberg scalar function whereas $\phi^{0} = f(t)$ and $\partial\phi^{0}/\partial t=\dot{f}(t)$ \cite{ArkaniHamed:2002sp}, and unitary gauge corresponds to the choice $f(t) = t$. To obtain the equations of motion, it is better to write eq. (\ref{1}) as a minisuperspace action
\begin{widetext}
\begin{eqnarray}\label{6}
\frac{\mathcal{S}}{V} =&& \int dt\Bigg\lbrace \PlanckMass^{2}\bigg[-3\frac{a\dot{a}^{2}}{N}-\Lambda a^{3}N\bigg]+\frac{\omega a^{3}}{2N}\dot{\sigma}^{2}\bigg[1-\frac{\beta}{N}\big(\frac{\dot{N}}{N^{2}}-H \big)\dot{\sigma}\bigg]
\nonumber\\&&
 +\PlanckMass^{2}m^{2}_{g}\bigg[Na^{3}(X -1)\big[3(X-2)-(X-4)(X-1)\alpha_3-(X-1)^{2}\alpha_{4}\big]
 \nonumber\\&& +f^{\prime} a^{4}X\bigg((X-1)\big[3-3(X-1)\alpha_{3}+(X-1)^{2}\alpha_{4}\big]\bigg)\bigg]\Bigg\rbrace ,
\end{eqnarray}
\end{widetext}
where $V$ is the comoving volume and we used the following definition
\begin{eqnarray}
\label{7}
X \equiv \frac{e^{\frac{\sigma}{\PlanckMass}}}{a}.
\end{eqnarray}
Note that $X$ is the ratio of scale factors of the metrics $e^{2\sigma/\PlanckMass}\tilde{g}_{\mu\nu}$ and $g_{\mu\nu}$. In addition, to further simplify the expressions, we define
\begin{eqnarray}\label{8}
H\equiv \frac{\dot{a}}{Na},
\end{eqnarray}
where $H$ is the Hubble parameter for the physical metric and
\begin{eqnarray}
\label{9}
r\equiv \frac{a}{N},
\end{eqnarray}
 corresponding to the ratio of the speeds of light on these two metrics. Note that we use integration by parts to obtain minisuperspace action (\ref{6}), so that we convert the second derivative terms into the first order derivatives as
 follows
\begin{eqnarray}\label{10}
\ddot{\sigma} \rightarrow -\frac{\dot{a}}{a}+\frac{\dot{N}}{N}\dot{\sigma}.
\end{eqnarray}
By varying the minisuperspace action (\ref{6}) with respect to $f$ and applying unitary gauge condition $f(t)=t$, we have
\begin{widetext}
\begin{eqnarray}\label{11}
\frac{d}{dt}\bigg \lbrace a^{4}(X-1)X\big[3+3\alpha_{3}+\alpha_{4}-(3\alpha_{3}+2\alpha_{4})X+\alpha_{4}X^{2}\big]\bigg\rbrace = 0.
\end{eqnarray}
Integrating the above equation yields
\begin{eqnarray}\label{12}(X-1)X\bigg[3+3\alpha_{3}+\alpha_{4}-(3\alpha_{3}+2\alpha_{4})X+\alpha_{4}X^{2}\bigg]=\frac{1}{a^{4}}\times C_{0},
\end{eqnarray}
\end{widetext}
where $C_{0}$ is an integral constant. It is worth mentioning that the constant solutions of $X$ lead to the effective energy density{\color{blue}{,}} and shows similar behavior to a cosmological constant. In an expanding Universe, the scale factor grows as time passes, so the right-hand side of that eq. (\ref{12}) decreases. Therefore, after a long enough time, $X$ leads to a constant value $X_{\rm SA}$ which makes the left-hand side of the equation zero.
One of the solutions for the eq. (\ref{12}) is $X = 0$ which leads to $\sigma\rightarrow 0$. Meanwhile, this solution multiplies to the perturbations of the auxiliary scalars, which means that we encounter strong coupling in the vector and scalar sectors. Thus, to avoid strong coupling, we discard this solution \cite{DAmico:2012hia}. Therefore, we have
\begin{widetext}
\begin{eqnarray} \label{13}(X-1)X\bigg[3+3\alpha_{3}+\alpha_{4}-(3\alpha_{3}+2\alpha_{4})X+\alpha_{4}X^{2}\bigg]\bigg |_{X=X_{\rm SA}}=0.
\end{eqnarray}
\end{widetext}
Another solution is $X = 1${\color{blue}{,}} which leads to a vanishing cosmological constant and because of inconsistency, it is unacceptable and so it should be ignored too \cite{DAmico:2012hia}. As a result, the two remaining solutions of the eq. (\ref{13}) are
\begin{eqnarray} \label{14}
X_{\rm SA}^{\pm}=\frac{3\alpha_{3}+2\alpha_{4}\pm \sqrt{9\alpha_{3}^{2}-12\alpha_{4}}}{2\alpha_{4}}.
\end{eqnarray}
Using this result{\color{blue}{,}}  we can find the modified Friedmann equation. For this purpose, we start by  writing the equation of motion related to the lapse function $N$,
\begin{widetext}
\begin{eqnarray} \label{15}
3H^{2}-\frac{\omega}{2\PlanckMass^{2}}\frac{\dot{\sigma}^{2}}{N^{2}}\big(1-3\beta \frac{\ddot{\sigma}}{N^{2}}\big) = \Lambda +m^{2}_{g}\bigg\lbrace(\alpha_{3}+\alpha_{4})X^{3}-3(1+2\alpha_{3}+\alpha_{4})X^{2}\nonumber\\+3(3+3\alpha_{3}+\alpha_{4})X-(6+4\alpha_{3}+\alpha_{4})\bigg\rbrace,
\end{eqnarray}
\end{widetext}
where $\frac{\dot{\sigma}}{N}=\PlanckMass\big(H+\frac{\dot{X}}{NX}\big)$ and $\frac{\ddot{\sigma}}{N^{2}}=\frac{\PlanckMass}{N^{2}}\frac{d}{dt}\big(NH+\frac{\dot{X}}{X}\big)$. If $X=X_{\rm SA}^{\pm}$, using time re-parametrization invariance to set $N(t) = 1$, we obtain the Friedmann equation
\begin{eqnarray} \label{16}
\bigg[3-\frac{\omega}{2}\big(1-3\beta \PlanckMass\dot{H}\big)\bigg]H^{2}=\Lambda +\xi^{\pm}.
\end{eqnarray}
Here $\xi^{\pm}$ is a fixed quantity, which is defined as follows
\begin{eqnarray} \label{17}
\xi^{\pm}=&& m^{2}_{g}\bigg\lbrace(\alpha_{3}+\alpha_{4}) (X_{\rm SA}^{\pm})^{3}-3(1+2\alpha_{3}+\alpha_{4})(X_{\rm SA}^{\pm})^{2} \nonumber\\
 &&+3(3+3\alpha_{3}+\alpha_{4})(X_{\rm SA}^{\pm})^{3}-(6+4\alpha_{3}+\alpha_{4})\bigg\rbrace.
\end{eqnarray}
One can see that $\xi^{\pm}$ has appeared as an additional cosmological constant in Friedmann's equation. For describing the dynamics of the Hubble parameter, it is better to convert time in eq. (\ref{16}) to the redshift. Applying  change of variable $\frac{d}{dt}=-H(z+1)\frac{d}{dz}$, the modified Friedmann equation becomes
\begin{eqnarray} \label{18}
\bigg[3-\frac{\omega}{2}\big(1+3\beta \PlanckMass H(1+z)H^{\prime}\big)\bigg]H^{2}=\Lambda +\xi^{\pm},
\end{eqnarray}
where prime denotes the derivative with respect to the redshift $z$. For the case $\beta=0$, the Friedmann equation (\ref{18}) provides a condition on the parameter $\omega$. For the self-accelerating solutions, to keep the left hand side of the Friedmann equation (\ref{18}) positive, one needs to have $\omega<6$.
This ensures having standard cosmology during matter domination if we add the ordinary matter to the right-hand side of the equation. For the case $\beta\neq0$, if $\omega$ is equal to $6$, the solution to the asymptotic state of eq. (\ref{18}) in small redshifts can be obtained as
\begin{widetext}
\begin{eqnarray} \label{19}
H(z) \sim H_{0}+\frac{\zeta^{\pm}}{H_{0}^{3}}z+\frac{ \left[-3 (\zeta^{\pm})^{2} -\zeta^{\pm} H_{0}^{4}\right]}{2 H_{0}^{7}}z^{2}+\frac{\zeta^{\pm}  \left[21 (\zeta^{\pm})^{2}+9 \zeta^{\pm} H_{0}^{4}+2 H_{0}^{8}\right]}{6 H_{0}^{11}}z^{3}+...
\end{eqnarray}
\end{widetext}
where $\zeta^{\pm}$ is defined as
\begin{eqnarray} \label{20}
\zeta^{\pm}\equiv -\frac{2}{9}\frac{\Lambda +\xi^{\pm}}{\beta \PlanckMass},
\end{eqnarray}
and $H_{0}$ represents the Hubble parameter at the present time. In the case $\omega = 6$, one can conclude that for any value of $\beta$, the Hubble parameter is well behaved and so this model can explain the accelerated expansion of the Universe.

Taking the variation of action (\ref{6}) with respect to the scalar field, the equation of motion corresponding to $\sigma$ is obtained as
 \begin{widetext}
\begin{eqnarray} \label{21}
0 && = 3H\Big(H+\frac{\dot{X}}{NX}\Big)+\frac{1}{N}\frac{d}{dt}\Big(H+\frac{\dot{X}}{NX}\Big)+\frac{3}{2}\beta \Big(H+\frac{\dot{X}}{NX}\Big) \Xi  -\frac{m^{2}_{g}}{\omega}X \nonumber\\
&&\times \Big\lbrace -(3+r)(3+3\alpha_{3}
+\alpha_{4})+6(1+r)(1+2\alpha_{3}+\alpha_{4})X-3(1+3r)(\alpha_{3}+\alpha_{4})X^{2}+4r\alpha_{4}X\Big\rbrace,\nonumber\\
 \end{eqnarray}
where
\begin{eqnarray} \label{22}
\Xi \equiv && \PlanckMass\bigg\lbrace 3H^{2}+\frac{\dot{H}}{N}+\frac{\dot{N}^{2}}{N^{4}}-5H\frac{\dot{N}}{N^{2}}-\frac{\ddot{N}}{N^{3}}\bigg\rbrace \Big(H+\frac{\dot{X}}{NX}\Big)\nonumber\\
&&+2\PlanckMass\Big(H-\frac{\dot{N}}{N^{2}}\Big)\frac{1}{N^{2}}\frac{d}{dt}\Big(NH+\frac{\dot{X}}{X}\Big).
\end{eqnarray}
Exploiting  eq. (\ref{21}), one finds
\begin{align}\label{rSA}
r_{\rm SA} &= \frac{1}{m_{g}^{2}(\omega - 6)(X_{\rm SA}^{\pm})^2(X_{\rm SA}^{\pm}\alpha_{3}-\alpha_{3}-2)}\bigg\lbrace 2\Big(m_{g}^{2}(3+\alpha_{3})-\Lambda\Big)\omega -2\beta (\omega -6)\omega H^{4}\PlanckMass^{2} \nonumber\\
& +m_{g}^{2}X_{\rm SA}^{\pm}\bigg[X_{\rm SA}^{\pm}\big(12 +6\alpha_{3}+4\omega +5\alpha_{3}\omega - X_{\rm SA}^{\pm}\alpha_{3}(\omega +6)\big)-6\omega (\alpha_{3}+2) \bigg]\bigg\rbrace.
\end{align}

\end{widetext}
\section{Cosmological tests}\label{C-t}

\begin{figure}
\centering
\includegraphics[width=6cm]{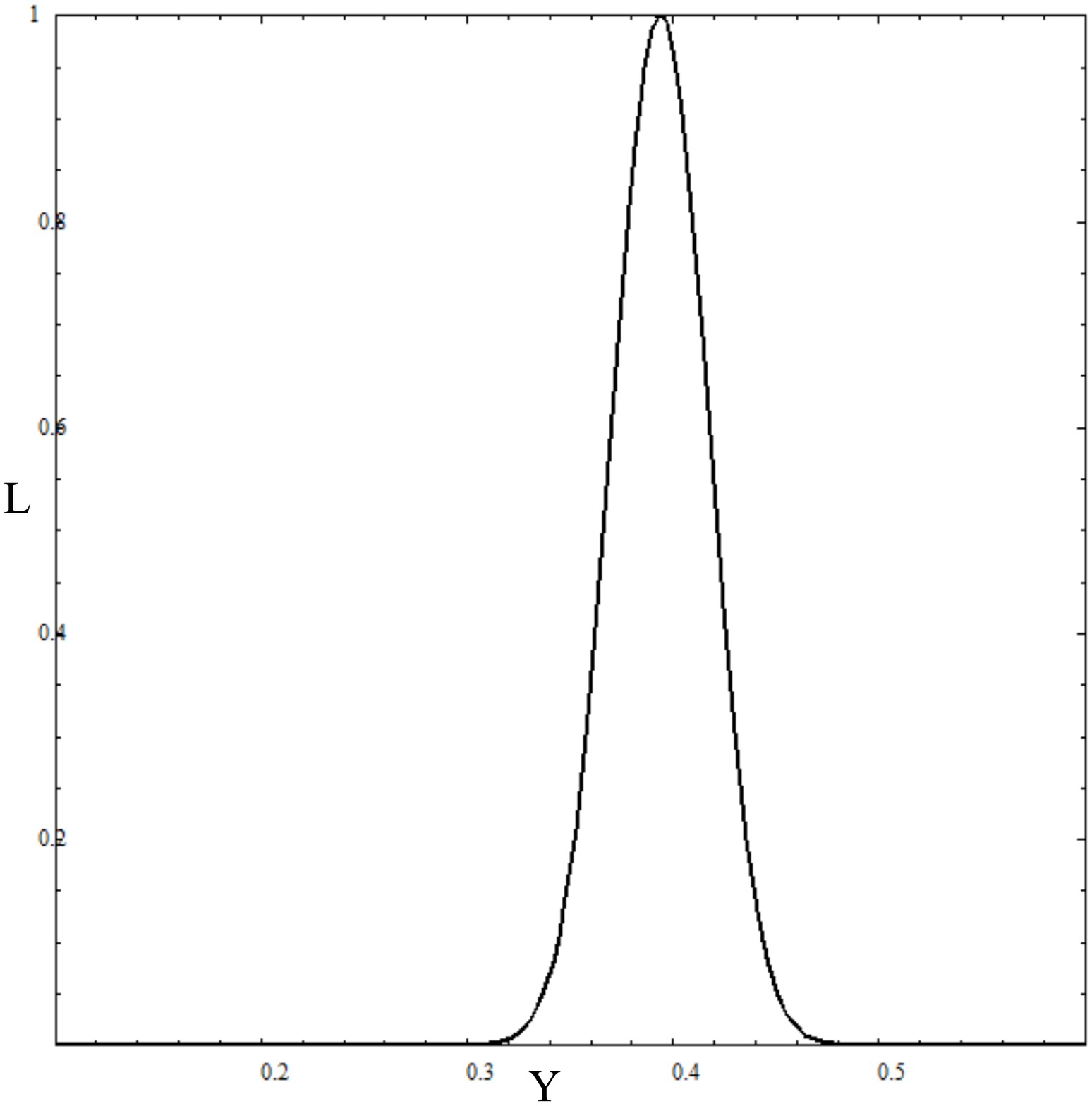}
\includegraphics[width=6cm]{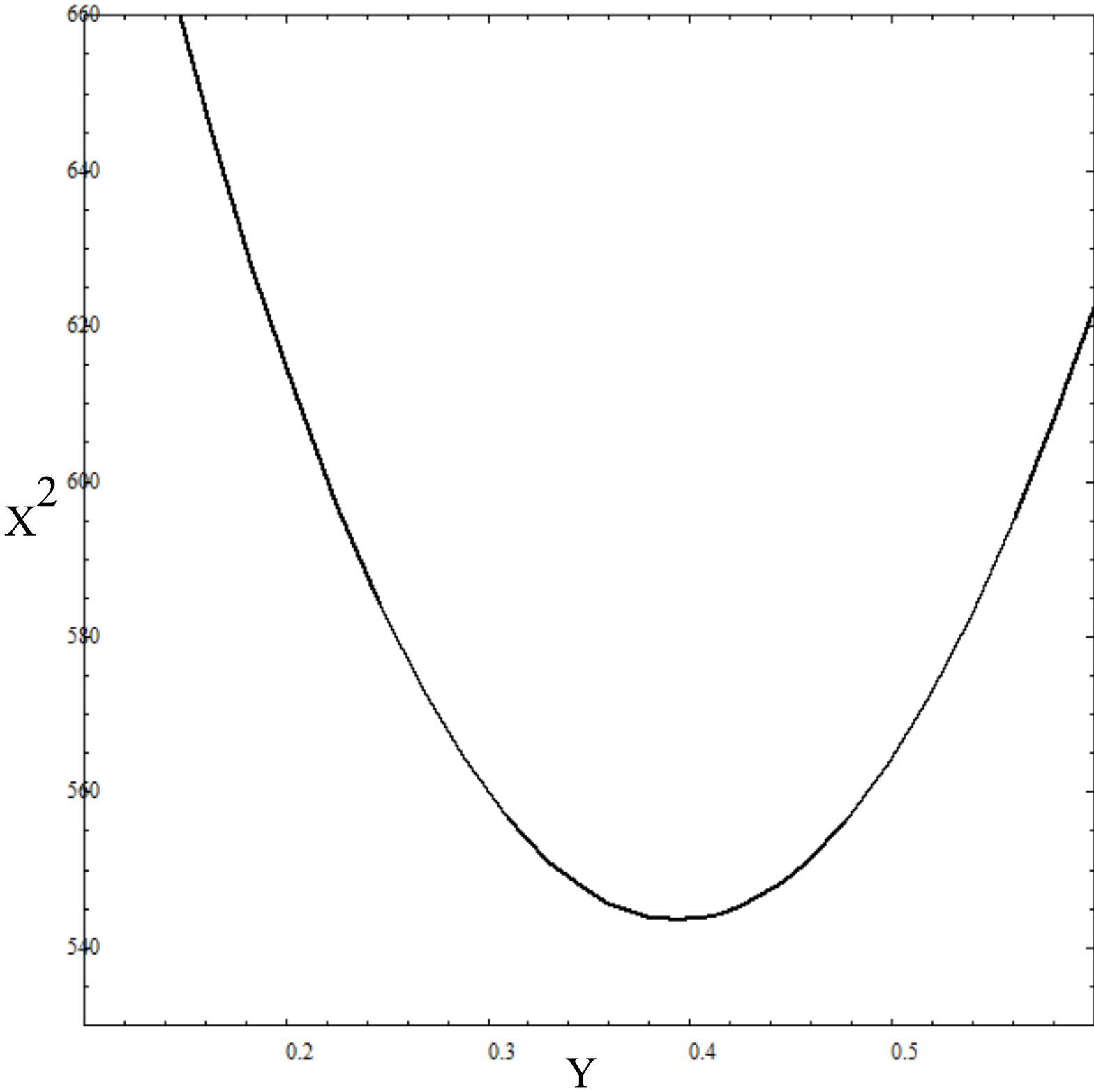}
\caption[figs]
{The $X^{2}$ and likelihood as functions of parameter $Y$.}\label{fig:H1}
\end{figure}

\begin{figure}
\centering
\includegraphics[width=7cm]{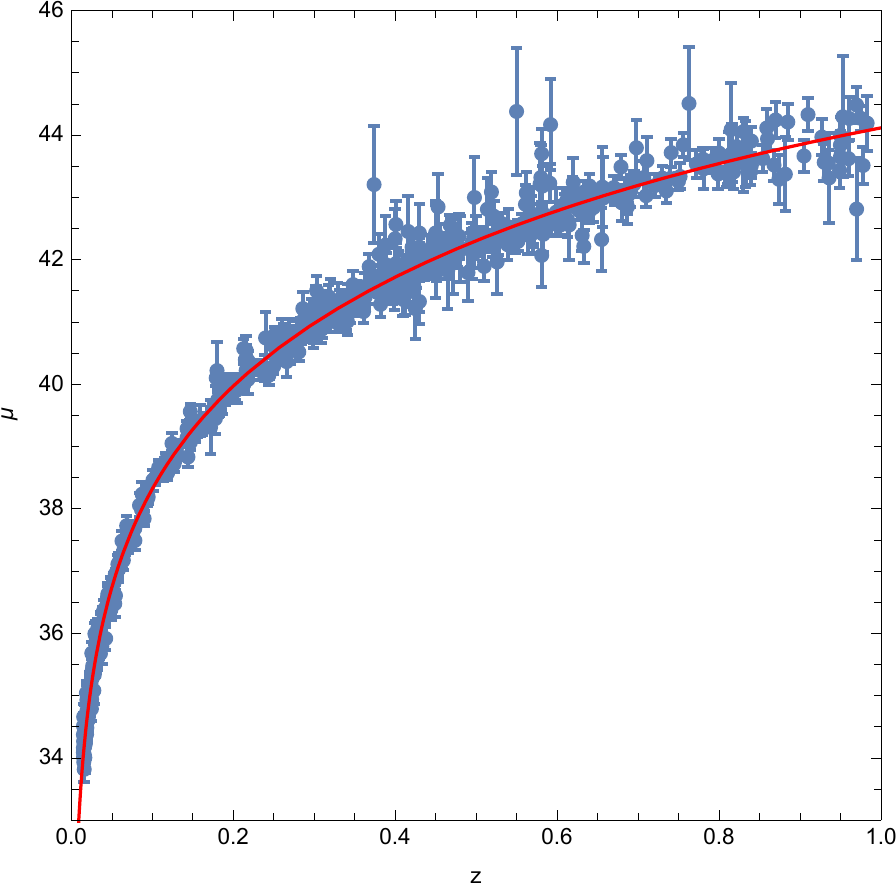}
\caption[figs]
{The Hubble diagram for the best fit (red solid line), comparing with the 557 Union2 SNIa data points (blue dots).}\label{fig:HP1}
\end{figure}

In 1998, observations on distant type Ia supernovae confirmed the accelerated expansion of the Universe \cite{Copeland:2006wr,Frieman:2008sn,Perlmutter:2003kf,Yang:2019fjt}. In this section, using the Union2 SNIa dataset consisting of 557 SNIa \cite{Amanullah:2010vv}, we examine the cubic Galileon massive gravity model. The results of the Union2 SNIa dataset can be expressed in terms of $\mu_{\rm obs}$, and should be compared with the predictions of the model
\begin{eqnarray}
\mu_{\rm th}(z_{i})=5\log_{10}D_{L}(z_{i})+\mu_{0},
\end{eqnarray}
where $\mu_{0}=42.38-5\log_{10}h$ ($h$ is the Hubble constant $H_{0}$ in units of $100 \, {\rm km/s/Mpc}$){\color{blue}{,}} and
\begin{eqnarray}
D_{L}(z)=(1+z)\int_{0}^{z}\frac{dx}{E(x;p)},
\end{eqnarray}
where $E=H/H_{0}$ and $p$ denotes the model parameters.
It should be noted that $X^{2}$ from the 557 Union2 SNIa is given by
\begin{equation}\label{3.3}
X_{\mu}^{2}(p)=\sum_{i}\frac{[\mu_{\rm obs}(z_{i})-\mu_{\rm th}(z_{i})]^{2}}{\sigma^{2}(z_{i})},
\end{equation}
where $\sigma$ is related to $1\sigma$ error and the parameter $\mu_{0}$ is a nuisance parameter and is independent of the data points. According to Refs.~\cite{Nesseris:2005ur,DiPietro:2002cz}, we expand $X_{\mu}^{2}$ in eq. (\ref{3.3}) to minimize it with respect to $\mu_{0}$,
\begin{equation}\label{3.4}
X_{\mu}^{2}(p)=\tilde{A}-2\mu_{0}\tilde{B}+\mu_{0}^{2}\tilde{C},
\end{equation}
where
\begin{eqnarray}
&&\tilde{A}(p)=\sum_{i}\frac{[\mu_{\rm obs}(z_{i})-\mu_{\rm th}(z_{i};\mu_{0}=0,p)]^{2}}{\sigma_{\mu_{\rm obs}}^{2}(z_{i})}, \nonumber\\
&&\tilde{B}(p)=\sum_{i}\frac{\mu_{\rm obs}(z_{i})-\mu_{\rm th}(z_{i};\mu_{0}=0,p)}{\sigma_{\mu_{\rm obs}}^{2}(z_{i})}, \nonumber\\
&&\tilde{C}=\sum_{i}\frac{1}{\sigma_{\mu_{\rm obs}}^{2}(z_{i})}.
\end{eqnarray}
For $\mu_{0}=\frac{\tilde{B}}{\tilde{C}}$, eq. (\ref{3.4}) has a minimum at
\begin{equation}
\tilde{X}_{\mu}^{2}(p)=\tilde{A}(p)-\frac{\tilde{B}^{2}(p)}{\tilde{C}}.
\end{equation}
Since $X_{\mu, {\rm min }}^{2}=\tilde{X}_{\mu, {\rm min}}^{2}$, we can consider minimizing $\tilde{X}_{\mu}^{2}$ which is independent of $\mu_{0}$. It is important to note that the best-fit model parameters are determined by minimizing $X^{2}=\tilde{X}_{\mu}^{2}$. Clearly, the corresponding $h$ can be determined by $\mu_{0}=\frac{\tilde{B}}{\tilde{C}}$ for the best-fit parameters.
From eq. (\ref{19}), the dimensionless Hubble parameter can be written as
\begin{widetext}
\begin{eqnarray}
E=\frac{H(z)}{H_{0}}= 1+Yz-\bigg(\frac{3Y^{2}+Y}{2}\bigg)z^{2}+\bigg(\frac{21Y^{3}+9Y^{2}+2Y}{6}\bigg)z^{3}+...,
\end{eqnarray}
\end{widetext}
where
\begin{equation}
Y=\frac{\zeta^{\pm}}{H_{0}^{4}}.
\end{equation}
We plot the corresponding $X^{2}$ and likelihood as functions of parameter $Y$ in Fig.~\ref{fig:H1}. The best fit has $X_{\rm min}^{2}=543.579$, and the best-fit parameter is
\begin{align}
Y=& 0.394^{+0.147}_{-0.131}, \quad\quad  \mbox{with  1$\sigma$  uncertainty}, \\
Y=& 0.394^{+0.312}_{-0.250}, \quad\quad \mbox{with 2$\sigma$ uncertainty}.
\end{align}
\newpage
Furthermore, the best fit for the Hubble parameter in this theory is $h=0.701$. In Fig.~\ref{fig:HP1}, we illustrate the Hubble diagram for the best fit in comparison with the 557 Union2 SNIa data points. One can see that the cubic Galileon massive gravity is well consistent with the 557 Union2 SNIa dataset.

\section{Tensor perturbations}\label{T-p}

In this section, we analyze tensor perturbation to calculate the dispersion relation of gravitational waves. The first step is to find the action in the second order of the perturbations. For this, we consider small fluctuations $\delta g_{\mu\nu}$, and expand the physical metric around a background solution $g_{\mu\nu}^{(0)}$,
\begin{equation}
g_{\mu\nu}=g_{\mu\nu}^{(0)}+\delta g_{\mu\nu}.
\end{equation}
We require the terms in quadratic order in $\delta g_{\mu\nu}$ to be kept.
It should be noted that we perform all calculations in the unitary gauge, so we do not worry about the form of gauge-invariant combinations. Here, we raise and lower the spatial indices on perturbations by $\delta^{ij}$ and $\delta_{ij}$. Moreover, we can write the  expanded action in the Fourier domain with plane waves, i.e., with replacements  $\vec{\nabla}^{2}\rightarrow -k^{2}$ and $d^{3}x\rightarrow d^{3}k$. Note that we take $N=1$, which means that the derivatives should be considered with respect to time in the following calculations.

Let us begin by considering tensor perturbations as
\begin{eqnarray}
\delta g_{ij}=a^{2}h_{ij}^{TT},
\end{eqnarray}
where
\begin{equation}
\partial^{i}h_{ij}=0, \qquad g^{ij}h_{ij}=0.
\end{equation}
The tensor perturbed action in the second order can be calculated for each part of the  action separately. Gravity part of the perturbed action in quadratic order is
\begin{widetext}
\begin{eqnarray}
S^{(2)}_{\rm gravity}=\frac{\PlanckMass^{2}}{8}\int d^{3}k \, dt \, a^{3}\bigg[\dot{h}_{ij}\dot{h}^{ij}-\Big(\frac{k^{2}}{a^{2}}+4\dot{H}+6H^{2}-2\Lambda \Big)h^{ij}h_{ij}\bigg].
\end{eqnarray}
\\
The second order piece of massive gravity sector of the perturbed action can be written as
\begin{eqnarray}
S^{(2)}_{\rm massive}=&& \frac{\PlanckMass^{2}}{8}\int d^{3}k \, dt \, a^{3}m_{g}^{2}\bigg[(\alpha_{3}+\alpha_{4})rX^{3}-(1+2\alpha_{3}+\alpha_{4})(1+3r)X^{2}\nonumber\\
&&+(3+3\alpha_{3}+\alpha_{4})(3+2r)X-2(6+4\alpha_{3}+\alpha_{4})\bigg]h^{ij}h_{ij}.
\end{eqnarray}
In addition, we write cubic Galileon part of the perturbed action in quadratic order
\begin{eqnarray}
S^{(2)}_{\rm cubic-Galileon}=-\frac{\PlanckMass^{2}}{8}\int d^{3}k \, dt \, a^{3}\bigg[\frac{\omega}{\PlanckMass^{2}}\dot{\sigma}^{2}\big(1-\beta\dot{\sigma}^{2}\big)h^{ij}h_{ij}\bigg].
\end{eqnarray}
Summing up the second order pecies of the perturbed actions $S^{(2)}_{\rm gravity}$, $S^{(2)}_{\rm massive}$ and $S^{(2)}_{\rm cubic-Galileon}$, we obtain the total action in second order for tensor perturbations
\begin{eqnarray}
S^{(2)}_{\rm total}=\frac{\PlanckMass^{2}}{8}\int d^{3}k \, dt \, a^{3}\bigg\lbrace \dot{h}^{ij}\dot{h}_{ij}-\Big(\frac{k^{2}}{a^{2}}+M_{\rm GW}^{2}\Big)h^{ij}h_{ij}\bigg\rbrace .
\end{eqnarray}
At this point, using eqs. (\ref{14}) and (\ref{rSA}) we calculate $\alpha_{3}$ and $\alpha_{4}$. Therefore, the dispersion relation of gravitational waves is obtained as
\begin{eqnarray}
M_{\rm GW}^{2}=4\dot{H}+6H^{2}+\frac{\omega}{\PlanckMass^{2}}\dot{\sigma}^{2}\big(1-\beta \dot{\sigma}^{2}\big)+\gamma,
\end{eqnarray}
where
\begin{eqnarray}
\gamma =&& \frac{1}{(X_{\rm SA}^{\pm}-1)\big[2\omega -4\omega X_{\rm SA}^{\pm}+(6+\omega)(X_{\rm SA}^{\pm})^2\big]} \times \nonumber \\
&& \bigg\lbrace 2\beta (\omega -6)\omega H^{4}\PlanckMass^{2}\Big[(X_{\rm SA}^{\pm}-3)X_{\rm SA}^{\pm}(r_{\rm SA}X_{\rm SA}^{\pm} -2)-2\Big] \nonumber\\
&&+X_{\rm SA}^{\pm}\Big[6(m_{g}^{2}-\Lambda)(\omega(r_{\rm SA}-1)-2)+X_{\rm SA}^{\pm}\Big(2\Lambda(\omega(r_{\rm SA}-1)-6)+12m_{g}^{2}(3+\omega -r_{\rm SA}\omega)\nonumber\\
&&+m_{g}^{2}X_{\rm SA}^{\pm} \big[6\omega (r_{\rm SA}-1)+X_{\rm SA}^{\pm}(6+6r_{\rm SA}+\omega -r_{\rm SA}\omega)-36 \big]\Big)\Big]\bigg\rbrace.
\end{eqnarray}
\end{widetext}
It is interesting to note that if the square of the mass of gravitational waves is positive, the stability of long-wavelength gravitational waves is guaranteed. However, if it is negative, it should be tachyonic. Hence, as the mass of the tachyon is of the order of the Hubble scale, the instability should take the age of the Universe to develop.

\section{Conclusion}\label{Con}

In this work we have presented the cubic Galileon massive gravity theory which is a development of de Rham-Gabadadze-Tolley (dRGT) massive gravity theory. We have introduced the action and have found the full set of equations of motion for a FLRW background. To explain the late-time acceleration of the Universe, we have analyzed the self-accelerating background solutions.

In addition,  we have tested the solution of cubic Galileon massive gravity theory with the latest Union2 SNIa dataset, which consists of 557 SNIa, and have demonstrated the compatibility of the model with the observational data. Therefore, this comparison with observational data for the late-time acceleration of the Universe can be very useful for checking the parameters of the the cubic Galileon massive gravity theory. We have illustrated the Hubble diagram for the best fit in comparison with the 557 Union2 SNIa data points and the best fit for the Hubble parameter in this theory is $h = 0.701$.

In the last section of this paper, for examining the mass of graviton in the framework of cubic Galileon massive gravity theory, we have presented a detailed analysis of tensor perturbation and have obtained the dispersion relation of gravitational waves.  We have studied the propagation of gravitational perturbation in the FLRW cosmology in the cubic Galileon massive gravity theory. This kind of analysis is crucial for probing the alternative gravity theories in the era of gravitational waves.



\acknowledgments
We are really grateful to Nishant Agarwal for his very useful notes and codes on tensor perturbations. ARA, YI and SK would like to thank A. Emir Gumrukcuolu and Hao Wei for useful discussions and comments.
LS was supported by the National Natural Science Foundation of China (Grants No. 11975027, No. 11991053, No. 11721303), the National SKA Program of China (No. 2020SKA0120300), and the Max Planck Partner Group Program funded by the Max Planck Society.




\end{document}